\documentclass{ws-mpla}

\begin{document}

\markboth{Rakhimov O.G.} {{Magnetized Particle Motion Around Black
Hole in Braneworld}}

\title{Magnetized Particle Motion Around Black Hole in Braneworld}

\author{\footnotesize Rakhimov O.G.}

\address{Institute of Nuclear Physics, Ulughbek, Tashkent 100214,
Uzbekistan \\ orahimov.81@mail.ru}

\maketitle

\begin{abstract}

We investigate the motion  of a magnetized particle orbiting
around a  black hole in braneworld placed  in  asymptotically
uniform magnetic field. The influence of brane parameter on
effective potential of the radial motion of magnetized spinning
particle around the braneworld black hole using Hamilton–Jacobi
formalism is studied. It is found that circular orbits for photons
and slowly moving particles may become stable near $r = 3M$. It
was argued that the radii of the innermost stable circular orbits
are sensitive on the change of brane parameter. Similar discussion
without Weil parameter has been considered by de Felice et all
in~Ref. \refcite{rs99,98}.

\end{abstract}

\ccode{PACS Nos.: 04.50.-h, 04.40.Dg, 97.60.Gb.}

\section{Introduction}

One of the exact solutions of gravitational field equations in the
braneworld was obtained by Dadhich et al~\cite{nkd00} (so called
DMPR solution). This solution is the analog of solution of
Reissner-Nordstr\"{o}m, which describes the gravitational field of
black hole with electric charge $Q$~\cite{chandra}.

Observational possibilities of testing the braneworld black hole
models at an astrophysical scale have intensively discussed in the
literature during the last years, for example through the
gravitational lensing~\cite {k06,lens,wald,lens1}, the motion of
test particles~\cite {i05} and the classical tests of general
relativity (perihelion precession, deflection of light and the
radar echo delay) in the Solar system~\cite{bhl08}. The motion of
charged test particle around black hole in braneworld in the
presence of external uniform magnetic field have been studied in
Ref.~\refcite{aa09}.

A braneworld corrections to the charged rotating black holes and
to the perturbations in the electromagnetic potential around black
holes are studied in Ref.~\refcite{m04,ag05,rc06}. In
Ref.~\refcite{af08} authors considered  the stellar magnetic field
configurations of relativistic stars with brane charge.

Here we consider motion of magnetized particle around non-rotating
DMPR black hole in braneworld and motion of magnetized particle in
axial symmetric gravitational field. Innermost stable circular
orbits~\cite{hartle,barak}, storage and release of energy of
particle have been considered taking account of magnetic coupling
parameter $\beta$~\cite{rs99} in Schwarzschild spacetime (i.e.
$Q^*=0$). Following the idea of the paper Ref.~\refcite{rs99,umf},
we assume that the black hole is immersed in the external uniform
magnetic field. We plan to study the influence of brane parameter
and magnetic field on stable circular orbits of magnetized
particle around compact object in braneworld. We obtain the
equation of motion of the particles using Hamilton-Jacobi
formalism~\cite{mtw}.

The paper is organized as follows. In section 2 we introduce the
scenario for our model, specifying the spacetime metric and the
external magnetic field. In section 3 we discuss the
electromagnetic interaction of a magnetized particle in a curved
spacetime. In section 4 we solve the radial equation, obtained by
means of the Hamilton–Jacobi formalism, in order to find the
admissible circular orbits. Section 5 is devoted to the study of
the inner stable circular orbits, with some numerical estimates
and possible astrophysical implications. Finally, we conclude our
results in section 6.

We use in this paper a system of units in which $c = 1$, a
space-like signature $(-,+,+,+)$ and a spherical coordinate system
$(t,r,\theta ,\varphi)$. Greek indices are taken to run from 0 to
3, Latin indices from 1 to 3 and we adopt the standard convention
for the summation over repeated indices. We will indicate vectors
with bold symbols ({\it e.g.} ${\boldsymbol B}$) .

\section{\label{sec:emfield} Electromagnetic Field Around Black Hole}

Consider a black hole immersed in external asymptotically uniform
magnetic field ${\bf{B}}$. A particle of mass $m$, carrying a
magnetic dipole moment $\mu$, is assumed to move around the black
hole, following a circular orbit at equatorial plane. The magnetic
field is taken perpendicular to the orbital plane.

%{\bf{\emph{The spacetime}}}
Spacetime metric of the DMPR black hole solution in braneworld
takes form \cite{nkd00}

\begin{equation}\label{metric}
 ds^2=-A^2dt^2+H^2dr^2+r^2d\theta^2+r^2\sin^2\varphi,
 \end{equation}
where
$$A^2=H^{-2}=\left(1-\frac{2M}{r}+\frac{Q^*}{r^2}\right),$$
$Q^*$ is the bulk tidal charge and $M$ is the total mass of the
central black hole. The polar axis was choose along the direction
of ${\bf{B}}$. So we may assume, the particle motion was happening
in the equatorial plane.

%\section{The Magnetic Field}

The exact solution for black hole in braneworld immersed in
external uniform magnetic field is considered in
Ref.~\refcite{aa09}. The potential of the electromagnetic field
around black hole in braneworld has the form
\begin{equation}\label{pot}
 A^{\mu}=\tilde{A^{\mu}}+a^{\mu},
 \end{equation}
where $a^{\mu}$ proportional to the angular momentum of black hole
$a$ as
\begin{equation}\label{pot}
a^{\mu}=\frac{\pi a {B}}{6M^2}\left(1,0,0,0\right).
 \end{equation}
Since we consider the nonrotating black hole in braneworld, the
second term of the expression (\ref{pot}) vanishes. In this paper
we will consider only first term of the potential of the
electromagnetic field around black hole in braneworld, which has
the following form
\begin{equation}\label{epotential}
A_\mu= \frac12\delta^\varphi_\mu B r^2\sin^2\theta.
\end{equation}

Nonvanishing components of the Faraday electromagnetic tensor
$F_{\mu\nu}=\partial_\mu A_\nu-
\partial_\nu A_\mu$ reads:
\begin{eqnarray}\label{ten el f}
&& F_{r\varphi}=B_0r\sin^2\theta, \nonumber \\
&& F_{\theta\varphi}=B_0r^2\sin\theta\cos\theta.
\end{eqnarray}

Expressions {\ref{ten el f}} is the tensors of electromagnetic
field. They are identical with the tensors of electromagnetical
field in Schwarzschild metric~\cite{rs99}.

\section{The equation of motion}

The Hamilton-Jacobi equation~\cite{rs99}
\begin{equation}
\label{Ham-Jac-eq} g^{\mu\nu}\left(\frac{\partial S}{\partial
x^\mu}-qA_{\mu}\right)\left(\frac{\partial S}{\partial
x^\nu}-qA_{\nu}\right)=-m^2+mD^{\mu\nu}F_{\mu\nu}\ ,
\end{equation}
for motion of the charged test particles with mass $m$  is
applicable as a useful computational tool only when separation of
variables can be effected. Equation (\ref{Ham-Jac-eq})  can be
considered as the Hamilton-Jacobi equation for the particle,
interacting with external electromagnetic field~\cite{rs99}.
$D^{\mu \nu}$ is polarization tensor~\cite{rs99} and taken to be
proportional to the particle spin,
\begin{equation}\label{polar}
D^{\mu\nu}=\frac{q}{m}S^{\mu\nu},
\end{equation}
where $S^{\mu\nu}$ is the antisymmetric spin
tensor~\cite{rs99,97}.

Consider now particle  with zero charge ($q=0$) preserving a
magnetic dipole moment $\vec{\mu}$, due to some inertial
electrodynamical structure.

Then the  Hamilton-Jacobi equation (\ref{Ham-Jac-eq}) can be read
as:
\begin{equation}\label{motion eq}
H^2{\cal E}^2+\frac{1}{r^2\sin^2\theta}{\cal L}^2 + A^2
\left(\frac{\partial S_{\rm r\theta}}{\partial r}\right)
+\frac{1}{r^2}\frac{\partial S_{\rm r\theta}}{\partial
\theta}=-m^2 + 2m\mu B_0 K[\lambda_{\hat{\alpha}}].
\end{equation}
%
% \section{The electromagnetic interaction}

The magnetized particle interacts with the exterior magnetic field
$B$ via the $D \cdot F$ term in (\ref{motion eq}). In the present
case $D^{\mu\nu}$ will be proportional only to
$\overrightarrow{\mu}$. So, using the similarity with the spin
theory one can define  define~\cite{rs99}:
\begin{equation}\label{polar levi}
D^{\mu\nu}=\eta^{\mu\nu\rho\lambda}u_\rho\mu_\lambda,
\end{equation}
where $u^{\mu}$ is the 4-velocity and $\mu^{\lambda}$ is the
particle magnetic moment 4-vector. In our coordinate $D^{\mu\nu}$
is not constant. Such an equation has to be found and solved along
with (\ref{motion eq}). An alternative approach is to exploit the
fact that the interaction term $D \cdot F$ is a
scalar~\cite{rs99}.

From equation  (\ref{polar levi}) one can rewrite in the rest
frame of a fiducial comoving observer $u_{\nu}D^{\mu\nu}$ is
totally transverse, namely~\cite{rs99}
\begin{equation}\label{polar velo}
D^{\mu\nu}u_\nu=0.
\end{equation}

$F^{\mu\nu}$ posses decomposition
\begin{equation}
F_{\mu\nu}=-\eta_{\mu\nu\alpha\beta}B^\alpha u^\beta+u_\mu E_\nu-
u_\nu E_\mu.
\end{equation}
Here $B$ and $E$ are the magnetic and electric fields,
respectively. Admitting with (\ref{polar levi}) and using
(\ref{polar velo}) one can write for~\cite{rs99}
$D^{\mu\nu}F_{\mu\nu}$
\begin{equation}
D^{\mu\nu}F_{\mu\nu}=2\mu^\alpha B_\alpha.
\end{equation}

Finally, taking into account (\ref{ten el f}) and (\ref{polar
levi}), we may factorize a constant quantity out of the
interaction term $D \cdot F$, as~\cite{rs99}:
\begin{equation}
D\cdot F=2\mu B_0K[\lambda_{\hat{\alpha}}].
\end{equation}

\section{The circular orbits }

Our goal in this section is to explore orbital motion around the
compact object in braneworld, that is influences brane parameter
to the radii of stable orbits using  equation (\ref{motion eq}).
For an equatorial orbit one can write $\theta =\pi/2$ and
consequently we have $p_{\theta}=0$. The spacetime symmetries are
preserved by the axisymmetric configuration of the magnetic field
and therefore they still allow for two conserved quantities:
$p_{\phi} =L$ and $p_t=-E$. These are energy and angular momentum
of particle respectively.

Equations of motion (\ref{motion eq})  take the following form:
\begin{eqnarray}\label{energy}
&& 2M \frac{dt}{d\sigma}=
\left(1-\frac{1}{\rho}+\frac{{\hat{Q}}^*}{\rho^2}\right)^{-1}{e},
\end{eqnarray}
\begin{eqnarray}\label{eq mot}
 4M^2 \left(\frac{d\rho}{d\sigma}\right)^2= {e}^2-
V(\rho,\lambda,e,{\hat{Q}}^*),
\end{eqnarray}
\begin{eqnarray} \label{angular} 2M
\frac{d\varphi}{d\sigma}=\frac{\lambda}{\rho^2\sin^2\theta},
\end{eqnarray}
here ${\hat{Q}}^*=Q^*/4M^2$ and  $V(\rho,\lambda,e,{\hat{Q}}^*)$
is the effective potential and it was described in the following
form:
\begin{eqnarray}
V(\rho,\lambda,e,{\hat{Q}}^*)=\left(1-\frac{1}{\rho}+
\frac{{\hat{Q}}^*}{\rho^2}\right)
\left(1+\frac{\lambda^2}{\rho^2}-\beta
K[\lambda_{\hat{\alpha}}]\right).\nonumber
\end{eqnarray}

In equations (\ref{energy}), (\ref{eq mot}) and (\ref{angular})
$\sigma$ is the proper time of the particle. In these expressions
we introdused the following notations~\cite{rs99}:
\begin{eqnarray}
\rho=\frac{r}{2M}\nonumber,\qquad \qquad e=\frac{E}{m},
\qquad\qquad\nonumber \lambda=\frac{L}{2Mm},\qquad \qquad
\beta=\frac{2\mu B_0}{m},\nonumber
\end{eqnarray}
and
\begin{equation}\label{K}
K[\lambda_{\hat{\alpha}}]=h^{\psi}\left(1-\frac{1}{\rho}
+\frac{{\hat{Q}}^*}{\rho^2}\right)=h^{\psi} \frac{\Delta}{\rho^2},
\end{equation}
here
$h^{\psi}=(1-1/\rho+{\hat{Q}}^*/\rho^2-4M^2\Omega^2\rho^2)^{-\frac{1}{2}}$
~\cite{rs99}.

The parameter $\beta$ is responsible for the intensity of the
magnetic interaction~\cite{rs99}. In this paper we suppose
$\beta>0$.
And  now the circular orbits are obtained from
\begin{equation} \label{dif eff}
\frac{d \rho}{d \tau}=0,   \quad \quad \textrm{and}  \quad \quad
\frac{\partial V( \rho, e, \ell, \beta, {\hat{Q}}^* )}{\partial
\rho }=0.
\end{equation}
Using these equations we get
\begin{equation}\label{beta}
\beta(\rho,\lambda,e,{\hat{Q}}^*)=\frac{1}{K[\lambda_{\hat{\alpha}}]}\left[1-
\frac{\lambda^2}{\rho^2}-\frac{e^2\rho^2}{\rho^2-\rho+{\hat{Q}}^*}\right].
\end{equation}
Moreover, using equation  (\ref{eq mot}) and (\ref{dif eff}) one
can write for the first  derivative of potential~\cite{rs99}:
\begin{equation}\label{dif pot}
\frac{\partial V}{\partial
\rho}=\left(1-\frac{1}{\rho}+\frac{{\hat{Q}}^*}{\rho^2}\right)K[\lambda_{\hat{\alpha}}]
\frac{\partial \beta}{\partial \rho}.
\end{equation}
One can write  (\ref{beta}) and (\ref{dif pot}) in the following
form:
\begin{equation}\label{beta1}
\beta(\rho,\lambda,e,{\hat{Q}}^*)=\frac{\left(1-\frac{1}{\rho}+\frac{{\hat{Q}}^*}{\rho^2}
-4M^2\Omega^2\rho^2\right)^{\frac{1}{2}}}{\left(1-\frac{1}{\rho}
+\frac{{\hat{Q}}^*}{\rho^2}\right)}
\left(1+\frac{\lambda^2}{\rho^2}-\frac{e^2\rho^2}{\rho^2-\rho+{\hat{Q}}^*}\right),
\end{equation}
and
\begin{equation}
\frac{\partial V}{\partial
\rho}=\left(1-\frac{1}{\rho}+\frac{{\hat{Q}}^*}{\rho^2}\right)^2\left(1-\frac{1}{\rho}+\frac{{\hat{Q}}^*}{\rho^2}
-4M^2\Omega^2\rho^2\right)^{-\frac{1}{2}} \frac{\partial
\beta}{\partial \rho},
\end{equation}
using the equation (\ref{K}) for  the angular velocity of
particles together with two constant of motion $E$ and  $L$,
\begin{equation}
\Omega=\frac{\lambda}{2Me}\frac{\Delta}{\rho^4},
\end{equation}
one can rewrite (\ref{beta1}) in following form:
\begin{equation}\label{beta2}
\beta(\rho,\lambda,e,{\hat{Q}}^*)=\left(\frac{\rho^2}{\rho^2-\rho+{\hat{Q}}^*}-\frac{\lambda^2}{e^2\rho^2}\right)^{\frac{1}{2}}
\left(1+\frac{\lambda^2}{\rho^2}-\frac{e^2\rho^2}{\rho^2-\rho+{\hat{Q}}^*}\right).
\end{equation}

Since  $h^{\psi}\geq 0$, $\partial V/\partial \rho$ and $\partial
\beta/\partial \rho$ have the same sign, thus, for a given value
of the parameter $\beta$, the determination of circular orbits as
well as the analysis of their stability reduces to the solution of
the following set of equations:
\begin{equation}\label{dif beta}
\beta =\beta(\rho,e, \ell,{\hat{Q}}^*), \quad \quad \quad
\frac{\partial \beta(\rho,e, \ell,{\hat{Q}}^*)}{\partial \rho}=0.
\end{equation}

This is a system of two equations with five unknowns $\beta$,
$\rho$, $e$, $\ell$,  and weil parameter ${\hat{Q}}^*$, so its
solutions can be parametrized in terms of any two of the five
independent variables. We will use as free parameters the magnetic
coupling. and the  orbital radius $\rho$. Our aim is then to find
the angular momentum $\lambda$ and the  energy $e$ of the particle
as functions of $\beta$ and $\rho$ for the innermost stable
circular orbits, i.e. stable orbits near the critical radius
$\rho_{crit} =3/2$ (orbit of photon).

\section{Towards inner circular orbits }

A computer analysis of equation $(\ref{dif beta})$ gives us
\begin{equation}
e_{min}(\rho,\lambda,{\hat{Q}}^*)=\frac{\sqrt{2} \lambda
(\rho^2-\rho+{\hat{Q}}^*)}{\sqrt{\rho^5-2 {\hat{Q}}^* \rho^4}}.
\end{equation}

Putting $e_{min}$ into (\ref{beta2}), we obtain
\begin{equation}
\beta_{min}(\rho,\lambda,{\hat{Q}}^*)=
\sqrt{\frac{\rho^2(2\rho^2-3\rho+4{\hat{Q}}^*)}{(\rho^2-\rho+{\hat{Q}}^*)^2}}\left(
\frac{\sqrt{2}}{2}-\frac{\lambda^2
(2\rho^2-3\rho+4{\hat{Q}}^*)}{\sqrt{2}\rho^2(\rho-2{\hat{Q}}^*)}\right).
\end{equation}
$\beta_{min}$ yields the value of the magnetic coupling $\beta$
for a circular stable orbit having radius $\rho$ with parameter
$\lambda$ and $e_{min}$.

The locus $\beta_{extr}(\rho,{\hat{Q}}^*)$ of the maxima of
$\beta_{min}$ is obtained by solving $\partial\beta_{min}/\partial
\rho = 0$ with respect to $\lambda$, and inserting the solution in
$\beta_{min}$ again. This gives
\begin{equation}\label{ext beta}
\beta_{extr}(\rho,\lambda,{\hat{Q}}^*)=
\sqrt{\frac{\rho^2(2\rho^2-3\rho+4{\hat{Q}}^*)}{(\rho^2-\rho+{\hat{Q}}^*)^2}}\left(\frac{\sqrt{2}}{2}+\frac{2\rho^2-3\rho+4{\hat{Q}}^*}
{\sqrt{2}(\rho^2-4{\hat{Q}}^*\rho+{\hat{Q}}^*)}\right).
\end{equation}
In equation (\ref{ext beta}) we neglect infinitely small
expressions (i.e ${\hat{Q}}^*\cdot\beta^2$) and  obtain maximum
value $\rho_{+}$ of $\rho$ for circular orbits
\begin{equation}
\rho_{+}=\frac{4\beta^2-27-3\sqrt{81-8\beta^2-288{\hat{Q}}^*}}{4(\beta^2-9)}.
\end{equation}

For any given value of $\beta$ we also obtain value $\rho_{-}$
from $\beta=\beta_{min}\mid_{\lambda=0}$. Actually, $\rho_{-}$ is
the minimum value of $\rho$ for admissible circular stable orbits,
\begin{equation}
\rho_{-}=\frac{2\beta^2-3-\sqrt{9-12\beta^2-32{\hat{Q}}^*}}{2(\beta^2-2)}.
\end{equation}
Thus for a given values of parameters $\beta$ and ${\hat{Q}}^*$,
stable circular orbits near $\rho_{crit}$ are confined in the
range
\begin{equation}\label{rang}
\rho_{-}(\beta,{\hat{Q}}^*)<\rho<\rho_{+}(\beta).
\end{equation}
Obviously $\beta\ll1$ and we may neglect $\beta^4$ and
$\beta^2{\hat{Q}}^* $, and obtain
\begin{equation}
\Delta\rho(\beta,{\hat{Q}}^*)=\rho_{+}-\rho_{-}=\frac{\beta^2(4\beta^2-29+56{\hat{Q}}^*)}{6(18-11\beta^2)}.
\end{equation}

Then make transformation $\rho=\frac{r}{2M}$ and $\beta=\frac{2\mu
B_0}{m}$ and take into consideration the particle with $\mu \simeq
\mu_{Bohr}=5.27\times 10^{-21}$ erg $G^{-1}$ (the Bohr magneton),
$mass\simeq m_{p}=1.67\times 10^{-24}$g (the proton rest mass) and
magnetic field $B_{0}\simeq 10^{12}$G  we obtain~\cite{rs99}
$\beta \simeq 7\times 10^{-6}$. Assumption $M\simeq 10^{3}
M_{\odot}$, we adduce this results in table1:

The dependence of the  values of $e_{geod}$ and $e_{min}$ from
brane parameter. Evidently from this table, the value of
$e_{geod}$ is increasing with increase the of module of brane
parameter, but the value of $e_{min}$ inversely proportional to
the brane parameter.

\begin{table}[h]
\tbl{The dependence of the $\rho_{prop}$ of brane parameter $Q$.
$Q=0$ corresponds to Schwarzschild black hole, which was
considered by de Felice at all. }
{\begin{tabular}{@{}cccccccc@{}} \toprule Q   & 0    &
$3\cdot10^{-5}$    & $1\cdot10^{-4}$  & $1\cdot10^{-3}$   &
$3\cdot10^{-3}$  & $1\cdot10^{-2}$ & $3\cdot10^{-2}$ \\
\colrule $r_{prop} $(m)   & 100    & 150    & 151  & 152   & 153 &
155 & 160  \\
\botrule
\end{tabular}\label{ta1} }
\end{table}

Let us now choose a circular stable orbit with a value of $\rho$
in the allowed range (\ref{rang}). We evaluate for such an orbit
the corresponding $e$ and $\lambda$. From
$\beta=\beta_{min}(\rho,\lambda,{\hat{Q}}^*)$ we obtain

\begin{eqnarray}\label{ang min}
\lambda_{min}(\rho,\beta,{\hat{Q}}^*)=
\sqrt{\frac{2\beta\rho^4(2{\hat{Q}}^*-\rho)
(\rho^2-\rho+{\hat{Q}}^*)
-\sqrt{2\rho^2(2\rho^2-3\rho+4{\hat{Q}}^*)})}{ 2(\rho^2(2\rho^2
-3\rho+4{\hat{Q}}^*))^{\frac{3}{2}}}}.
\end{eqnarray}

Also, from $e(\rho,\lambda,{\hat{Q}}^*)$ and using (\ref{ang
min})we obtain energy

\begin{eqnarray}\label{ener min}
e_{min}(\rho,\beta,{\hat{Q}}^*)=
2^{\frac{1}{2}}\sqrt{\frac{(\rho^2-\rho+{\hat{Q}}^*) \sqrt{
2\rho^2-3\rho+4 {\hat{Q}}^*}-2^{\frac{1}{2}}\beta(\rho^2
-\rho+{\hat{Q}}^*)^2}{\rho^3(2\rho^2-3\rho+
4{\hat{Q}}^*)^{\frac{3}{ 2}}}}.
\end{eqnarray}
Equations (\ref{ang min}) and (\ref{ener min}) yield the required
values for given $\beta$, $\rho$ and ${\hat{Q}}^*$.

 In table 1 we provide the
numerical values for shrink of stable orbits for photonic particle
for the typical values of brane parameter. Apparently, the shrink
of stable orbits depends on the module of brane parameter. Decreasing
of brane parameter leads to increasing of radius of stable orbits.
In the case when Weil parameter is equal to zero  we get the results obtained in the
Schwarzschild metric~\cite{rs99}.

\section{Storage and release of energy }

The most interesting result which stems from the previous analysis
is that the influence of brane tension  parameter to the magnetic
interaction, where provides a very efficient mechanism of
'particle confinement'~\cite{rs99}. As far as we know  the
combined effect of curvature and  the pronounced relativistic
character of the particle motion in the vicinity of the circular
photon orbit, causes an amplification of the magnetic field which,
in turn, provides not only the binding force required to sustain
the orbital motion but also a large amount of magnetic interaction
energy which adds to the gravitational one. So we end with a tiny
ring of stable orbits with a high content of 'negative' binding
energy (magnetic and gravitational) which is subtracted from the
positive (kinetic and rest) energy of the relativistic particle to
give $e_{min}(\rho,\beta,{\hat{Q}}^*)$. Following to [1] it is
instructive to compare $e_{min}(\rho,\beta,{\hat{Q}}^*)$ with the
corresponding energy $e_{geod}(\rho,{\hat{Q}}^*)$ of a geodesic
particle (no magnetic interaction: $\beta = 0$) moving on the same
orbit:
\begin{eqnarray}
e_{geod}(\rho,{\hat{Q}}^*)= \frac{\rho^2-\rho+{\hat{Q}}^*}{{\hat{
Q}}^*}\sqrt{\frac{2}{2\rho^2-3\rho+4{\hat{Q}}^*}}.
\end{eqnarray}
We emphasize that $e_{geod}$ is the energy of a geodesic unstable
orbit, while $e_{min}$ is the energy of a magnetized stable one.
Then the quantity
\begin{eqnarray}
\Delta e(\rho,\beta,{\hat{Q}}^*)= e_{min}(\tilde{\rho}, \beta,
{\hat{Q}}^*)-e_{geod}(\rho,{\hat{Q}}^*) =\nonumber\\
&&\hspace{-6.6cm}  -e_{geod}(\rho,{\hat{Q}}^*)
\left[1-\sqrt{1-\beta e_{geod}(\rho,{\hat{Q}}^*)}\right],
\end{eqnarray}
corresponds to the (negative) magnetic binding energy, stored in
that orbit. This would be just the maximum energy released in the
case of an abrupt disappearance of the magnetic coupling $\beta$.
In table 2 we expressed the dependence of brane parameter
${\hat{Q}}^*$ from $e_{geod}(\tilde{\rho},{\hat{Q}}^*)$ and
$e_{min}(\tilde{\rho},\beta,{\hat{Q}}^*)$.

Making the numerical calculation one can write some values for
$e_{geod}(\tilde{\rho},{\hat{Q}}^*)$ in the  following  table,
$\tilde{\rho}=\frac{1}{2}\simeq(\rho_{+}+\rho_{-})=
\rho_{crit}+\frac{13}{48}\beta^2-{\hat{Q}}^*$. \quad
$e_{geod}=e_{geod}(\tilde{\rho},{\hat{Q}}^*)$,\quad
 $e_{min}=e_{min}(\tilde{\rho},\beta,{\hat{Q}}^*)$ .
\begin{figure}[ph]
\caption{Radial dependence of effective potential $\beta$ for
constant values of $e$ and $\lambda$. The value of  ${\hat{Q}}^*$
is different for each graph.\protect\label{fig1}}
\centerline{\psfig{file=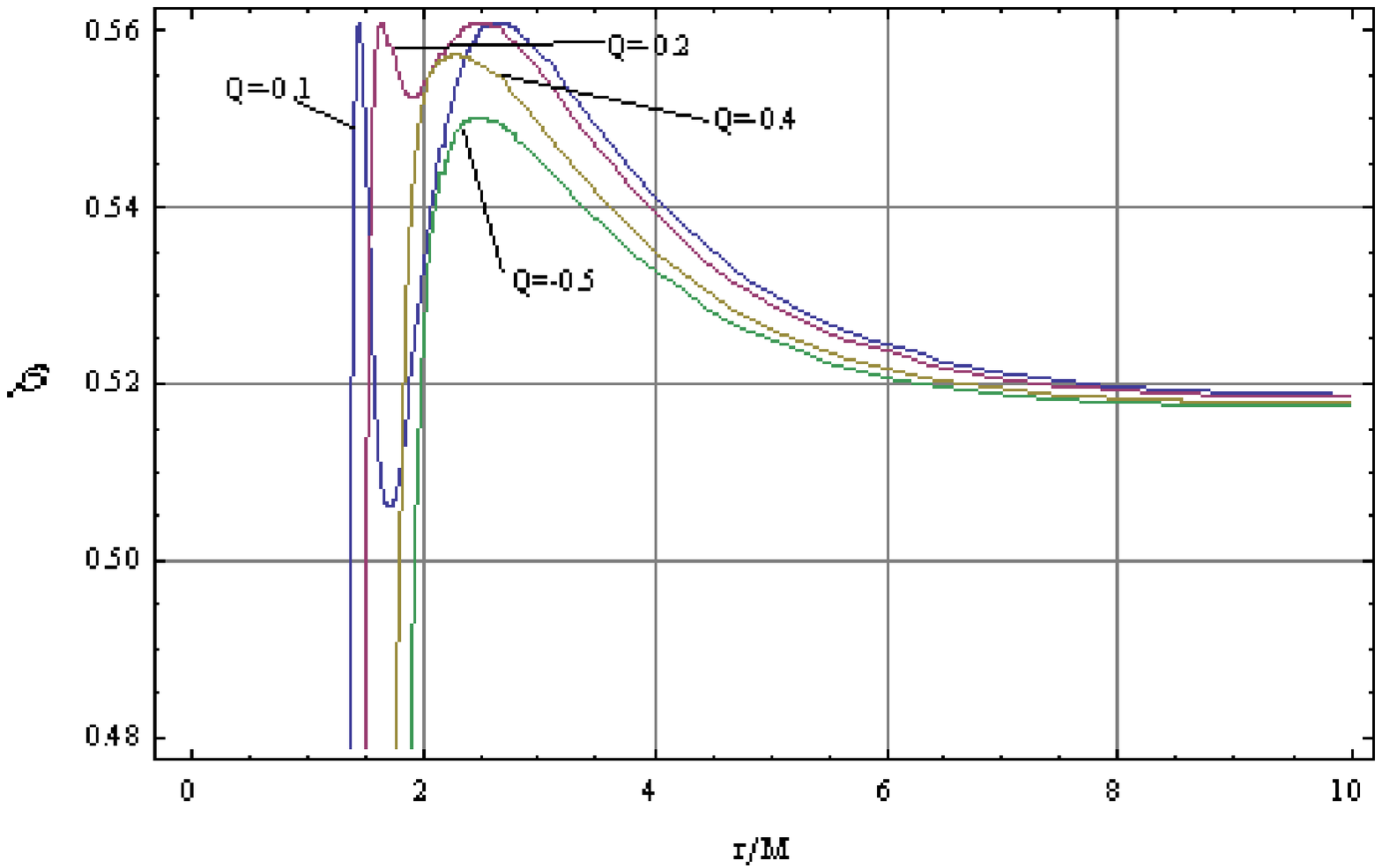,width=4.0in}} \vspace*{8pt}
\end{figure}

\begin{table}[h]
\tbl{The dependence of the  values of $e_{geod}$ and $e_{min}$
from brane parameter. Evidently from this table, the value of
$e_{geod}$ is increasing with increase the of module of brane
parameter, but the value of $e_{min}$ inversely proportional to
the brane parameter.}
{\begin{tabular}{@{}cccccccc@{}} \toprule Q   & 0    &
$1\cdot10^{-13}$ & $3\cdot10^{-13}$  & $1\cdot10^{-12}$   &
$3\cdot10^{-12}$  & $5\cdot10^{-12}$   & $7\cdot10^{-12}$ \\
\colrule $e_{geod}$   & $11.2\cdot10^{4}$     & $11.3\cdot10^{4}$
& $11.5\cdot10^{4}$ & $11.9\cdot10^{4}$   & $13.6\cdot10^{4}$
&$16.2\cdot10^{4}$  & $21.3\cdot10^{4}$  \\
\colrule $e_{min}$   & $5.2\cdot10^{4}$     & $5.1\cdot10^{4}$ &
$5.06\cdot10^{4}$ & $5.01\cdot10^{4}$ & $4.9\cdot10^{4}$
&$4.7\cdot10^{4}$ &$4.4\cdot10^{4}$\\ \botrule
\end{tabular}\label{ta1} }
\end{table}

In table 2 we provide the value of energy of geodesic
particle (where $\beta=0$) and minimum energy of
particle ($\beta=\beta_{min}$). Decreasing of the value of brane
parameter $Q$, leads to increasing of $e_{geod}$ and decreasing
of $e_{min}$.

\section{Conclusion}

Using the Hamilton-Jacobi formalism, we have analytically solved
the radial equation for the motion of a magnetized particle
orbiting  braneworld black hole surrounded by a strong magnetic
field. The case of the Schwarzschild black hole was extensively
discussed in~\cite{rs99}, for example magnetic coupling parameter
$\beta$, which is function of $\rho,\lambda$ and $e$. Consequently,
in Schwarzschild case, a sudden change in the value of the
parameter $\beta$ could cause an abrupt release of  stored
energy, perhaps providing a novel mechanism for jets or bursts.

In this article we perform the similar calculations in metric of
DMPR~\cite{nkd00} (Dadhich N et al). In limiting case (i.e $Q=0$)
our expressions match the expressions, which were calculated in
the Schwarzschild spacetime. We also recall that the particle
magnetic moment $\mu$ has been taken parallel to the external
magnetic field.

Obviously, from figure 1 one can see the existence region of
stable circular orbits shifts to about $2 M$, as far as we know
the circular orbits around the Schwarzschild black hole appear to
be in $3/2M$. This implies, the region of existence of stable
orbits shifts sideways to observer, i.e. moves away off the black
hole.

\section*{Acknowledgements}

The author thanks Bobomurat Ahmedov and Ahmadjon Abdujabbarov for
their  help, useful advices, editing the text and making important
corrections and comments. This research is supported in part by
the UzFFR (projects 1-10 and 11-10) and projects FA-F2-F079 and
FA-F2-F061 of the UzAS.

\end{document}